\documentclass{article}
\usepackage{PRIMEarxiv}
\usepackage[english]{babel}

\usepackage[table,xcdraw]{xcolor}
\usepackage{tabularx}
\usepackage{longtable}
\usepackage{authblk} 
\usepackage{xspace}
\usepackage{graphicx}
\usepackage[colorlinks=true, allcolors=blue]{hyperref}

\usepackage{amsmath}
\usepackage{amssymb}

\usepackage{xcolor}

\newcommand{\kx}[1]{\textcolor{black}{#1}}%by kexin

\newcommand{\yy}[1]{\textcolor{black}{#1}}
\usepackage{booktabs}
\title{
Personalized Transcranial Electrical Stimulation: A Review of Computational Modeling and Optimization
}

\author[1,2]{Mo Wang} %12250099@mail.sustech.edu.cn
\author[1]{Kexin Zheng}
\author[1]{Yingyue Xin}
\author[1]{Xiang Chen}
\author[1]{Yiling Liu}
\author[3,4]{Huichun Luo} %luohuichun2015@163.com
\author[1]{Jingsheng Tang}
\author[5]{Tifei Yuan}  % ytf0707@126.com
\author[2]{Hongkai Wen} %hongkai.wen@warwick.ac.uk
\author[6]{Pengfei Wei} %101014012@seu.edu.cn
\author[1,*]{Quanying Liu}    % liuqy@sustech.edu.cn

\affil[1]{Department of Biomedical Engineering, Southern University of Science and Technology, Shenzhen, China}
\affil[2]{Department of Computer Science, University of Warwick, Coventry CV4 7AL, United Kingdom}
\affil[3]{Department of Anesthesiology, Renji Hospital, Shanghai Jiao Tong University School of Medicine, Shanghai, China}
\affil[4]{Key Laboratory of Anesthesiology (Shanghai Jiao Tong University), Ministry of Education, Shanghai, China}
\affil[5]{Shanghai Key Laboratory of Psychotic Disorders, Brain Health Institute, National Center for Mental Disorders, Shanghai Mental Health Center, Shanghai Jiao Tong University School of Medicine and School of Psychology, Shanghai, China}
\affil[6]{School of Biological Science and Medical Engineering, State Key Laboratory of Digital Medicine, Southeast University, Nanjing, China}

\affil[*]{\textit{Corresponding author:} \href{mailto:liuqy@sustech.edu.cn}{liuqy@sustech.edu.cn}}

\begin{document}
\maketitle

\begin{abstract}
\textbf{Objective}. Personalized transcranial electrical stimulation (tES) has gained increasing attention due to the substantial inter-individual variability in brain anatomy and physiology. While previous reviews have discussed the physiological mechanisms and clinical applications of tES, there remains a critical gap in up-to-date syntheses focused on the computational modeling frameworks that enable individualized stimulation optimization. \textbf{Approach}. This review presents a comprehensive overview of recent advances in computational techniques supporting personalized tES. We systematically examine developments in forward modeling for simulating individualized electric fields, as well as inverse modeling approaches for optimizing stimulation parameters. We critically evaluate progress in head modeling pipelines, optimization algorithms, and the integration of multimodal brain data. %This review provides a comprehensive overview of recent advances in computational techniques supporting personalized tES. We systematically examine progress in forward modeling for simulating individualized electric fields, as well as inverse modeling approaches for optimizing stimulation strategies. Developments in head modeling, optimization algorithms, and the integration of multimodal brain data are critically evaluated.
\textbf{Main results}. Recent advances have substantially accelerated the construction of subject-specific head conductor models and expanded the landscape of optimization methods, including multi-objective optimization and brain network-informed optimization. These advances allow for dynamic and individualized stimulation planning, moving beyond empirical trial-and-error approaches.
\textbf{Significance}. By integrating the latest developments in computational modeling for personalized tES, this review highlights current challenges, emerging opportunities, and future directions for achieving precision neuromodulation in both research and clinical contexts. 

\end{abstract}

\linespread{1.5}

\section{Introduction}
Transcranial electrical stimulation (tES) is a non-invasive neuromodulation technique that delivers low-intensity electrical currents to the scalp with the objective of modulating brain activity. Compared to invasive techniques such as deep brain stimulation (DBS), tES offers great advantages in safety, accessibility, and cost-effectiveness~\cite{Krause2023TranscranialES,huang2017measurements}.
Accumulating evidence indicates that tES can modulate neural oscillations and cortical excitability, thereby influencing a range of cognitive functions and behaviors~\cite{liu2018immediate,Maiella2022SimultaneousTE,rufener2016transcranial, moliadze2019after}. 
As such, tES has been widely applied as both a research tool to probe causal brain-behavior relationships and a therapeutic intervention for neurological and psychiatric conditions~\cite{tavakoli2017transcranial,seo2019relation,solanki2021investigating, rasmussen2021high,simula2022transcranial, zhang2022alpha}.

\begin{figure}[thbp]  
	\centering
	\includegraphics[width=0.98\linewidth]{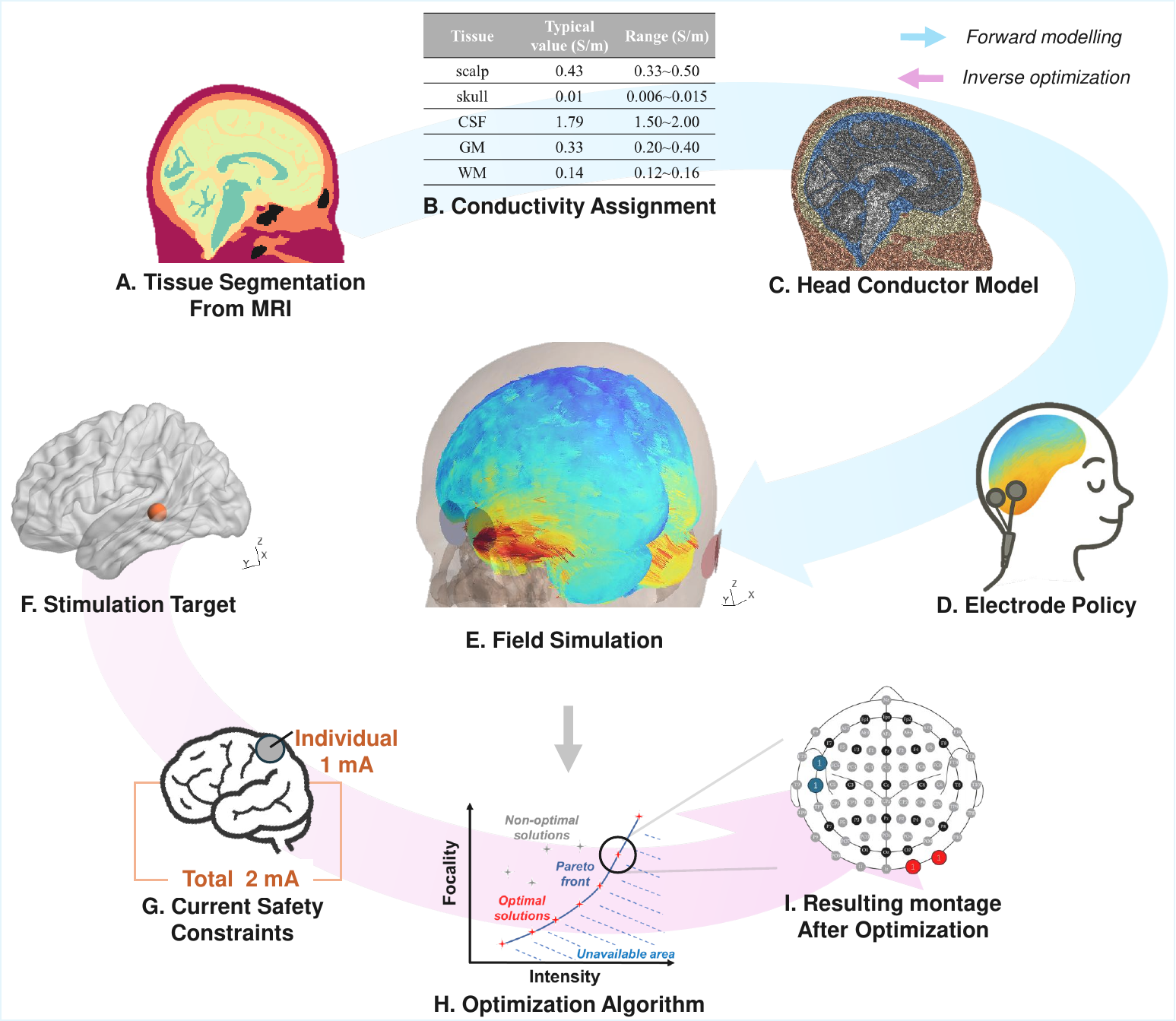}
	\caption{Overview of the forward modeling and inverse optimization for personalized tES.
(A) The construction of forward models for personalized computational simulations begins with tissue segmentation from MRI data.
(B) Conductivity values are assigned to each segmented tissue type to generate an individualized computational model.
(C) Head conductor model is generated by mesh generation and numerical methods.
(D) Virtual electrodes are then positioned on the model, and current is applied according to the predefined stimulation protocol.
(E) The electric field distribution is simulated by solving the current flow within the head model.
The stimulation montage is optimized based on the electric field simulation results from the forward model.
(F) The stimulation target is first identified, followed by (G) verifying current safety constraints.
(H) An optimization algorithm is then applied to the model to determine optimal stimulation parameters.
(I) Based on the optimization results, a stimulation montage is applied to the personalized tES for the subject. \yy{Steps (F–I) illustrate how the stimulation target, safety constraints, and optimization algorithms can use forward model results to determine optimal stimulation parameters. In this section, the focus remains on forward modeling, while inverse optimization is briefly illustrated to provide context on the overall computational workflow.}
	}
	\label{fig:framework}
\end{figure}

The efficacy of tES is influenced by various factors, including electric field strength~\cite{gaugain2023quasi,laakso2019can} and orientation~\cite{shahid2015use}, stimulation frequency, waveform characteristics~\cite{louviot2022transcranial}, and even the ongoing brain state~\cite{Krause2023TranscranialES,liu2018immediate,krause2019transcranial}.
Multiple tES variations have been developed, including transcranial direct current stimulation (tDCS)~\cite{Nitsche2000ExcitabilityCI,Brunoni2012ClinicalRW}, transcranial alternating current stimulation (tACS)~\cite{Antal2008ComparativelyWA,Bchinger2017ConcurrentTR}, transcranial random noise stimulation (tRNS)~\cite{vanderGroen2016TranscranialRN}, intersectional short pulse stimulation (ISP)~\cite{voroslakos2018direct}, and  transcranial temporal interference stimulation (tTIS)~\cite{grossman2017noninvasive,violante2022non,wang2022multi}.
Because each variation has distinct mechanisms of action and application potential, computational models are essential tools for their investigation.
Computational models in tES serve a dual role: forward simulation (Fig.\ref{fig:framework} A-E) and inverse optimization (Fig.\ref{fig:framework} F-I). Forward modeling, often referred to as head volume conduction modeling, predicts the spatial distribution of electric fields induced within the brain during stimulation. This capability is essential for mechanistically understanding how tES modulates neural activity~\cite{lee2021future}.
In early studies, researchers used template head models (e.g., MNI152 template) to simulate the electric field. While computationally efficient, such templates can overlook individual anatomical and physiological variability, reducing predictive accuracy. Individualized models, while demanding high-quality imaging and advanced segmentation, enable precise electric field estimation and tailored stimulation optimization.
Consequently, constructing subject-specific head models from individual MRI scans has become critical for achieving accurate simulations. 
Alongside the evolution of forward modeling techniques, tES hardware has also advanced—from simple two-electrode setups to complex multi-electrode arrays—substantially expanding the parameter space for stimulation configuration. Manual montage design or simple forward simulation is often insufficient to identify optimal stimulation strategies in this high-dimensional parameter space~\cite{hunold2023review}. Inverse modeling has thus emerged as an essential toolset: it has been widely applied in EEG and ECoG source localization to reconstruct neural current sources from measured scalp potentials, and it has been adapted to tES for optimizing electrode configurations~\cite{grech2008review,baillet2002electromagnetic,michel2004eeg}. By integrating personalized forward models with optimization algorithms, inverse modeling systematically determines individualized tES strategies that maximize efficacy, enhance target specificity, and minimize unintended effects.

A growing body of work demonstrates substantial inter-individual variability in neuroanatomy and neurophysiology, motivating increasing interest in personalized tES strategies~\cite{wang2022multi, violante2023non, ma2024mapping, grover2022long}. Although several prior articles discuss personalization in tES, most reviews primarily emphasize physiological mechanisms or clinical efficacy~\cite{liu2018immediate,yavari2018basic,chang2022application,gomez2024perspectives}. By contrast, surveys on computational modeling often provide limited, practice-oriented guidance for constructing forward models and public pipelines~\cite{gomez2024perspectives,berger2025human,bikson2012computational,van2023outcome,fernandez2020unification}; several omit optimization algorithms~\cite{hunold2023review,berger2025human,bikson2012computational,van2023outcome, bai2013review} and others do not cover tTIS or personalized modeling ~\cite{salvador2025group,miranda2018realistic}.

In this review, we place greater emphasis on the evolution of computational modeling itself, providing a systematic perspective on how personalized tES simulation and optimization can be implemented from the ground up. We further discuss potential enhancement techniques, including forward-model refinement through segmentation, conductivity, and anisotropy updates, neural-dynamics-integrated personalized modeling, and brain-state-informed inverse optimization.
Specifically, Section 2 focuses on forward modeling for predicting electric field distributions, while Section 3 examines inverse optimization for designing personalized stimulation strategies. These sections include 64 key studies: 22 modeling and theoretical works (e.g., head model construction, electric field computation, conductivity modeling), 34 studies on stimulation optimization (covering traditional algorithms, machine learning–based methods, and individualized electrode layout design), and 8 studies involving model validation or experimental verification. Section 4 introduces advanced techniques to enhance tES modeling, incorporating 17 studies with 12 proposed frameworks and 5 experimental validation. Section 5 examines current limitations and future perspectives, particularly in relation to the integration of neuroimaging techniques, and synthesizes insights from 13 recent innovative studies.
Finally, we highlight the current limitations of computational modeling in tES and outline potential directions for future research and applications.

\section{Forward modeling for personalized tES}

Forward modeling, also known as volume conduction modeling, plays a fundamental role in simulating the distribution of electric fields generated by tES (Fig.\ref{fig:framework} A-E). The earliest modeling work employed simplified spherical head models, which approximated the human head as three nested concentric spheres representing the scalp, skull, and brain, respectively. % Despite their simplicity, these models effectively captured key aspects of electric field distribution and demonstrated reasonable agreement with experimental measurements~\cite{miranda2006modeling}.
Despite their simplicity, these models effectively captured key aspects of electric field distribution, such as the spatial variation of field intensity across tissue layers and the general orientation of induced fields~\cite{rush2008eeg}, and demonstrated reasonable agreement with experimental measurements, including scalp electrode recordings and intracranial measurements in in vivo studies~\cite{miranda2006modeling}. As the field progressed and the demand for realistic head models increased, researchers began incorporating real anatomical data into head conductivity models using MRI-based imaging.
The construction of realistic forward models typically involves several major steps: tissue segmentation, mesh generation, and electric field simulation. In the early stages, constructing individualized models was time-consuming and labor-intensive, prompting the widespread use of standardized templates for forward modeling.
Among the most prominent templates is the MNI152 template head~\cite{evans19933d, fonov2011unbiased}, which represents the average of 152 T1-weighted anatomical scans. FEM-based TES models, such as the New York Head Model introduced by Huang et al.~\cite{huang2016new}, are built upon such anatomical templates and incorporate additional information, including tissue conductivity, to simulate electric fields. These resources have significantly advanced large-scale and reproducible tES research~\cite{asadzadeh2020systematic, salehinejad2020transcranial, zorzos2021advances}.
\yy{However, for normalized models, substantial inter-individual variability in head anatomy and tissue conductivity, which can lead to marked differences in electric field profiles even under identical stimulation conditions~\cite{kasten2019integrating,von2021interindividual,antonenko2021inter}. By contrast, individualized models are based on subject-specific MRI data, allowing accurate reconstruction of anatomical details and tissue properties. While they require high-quality imaging and advanced segmentation, these models enable subject-tailored optimization of stimulation parameters, thereby providing more precise electric field predictions and greater suitability for clinical applications.}
Recent advances in neuroimaging pipelines and high-performance computing have made the construction of subject-specific head models increasingly practical. Personalized head models can now be generated from individual MRI within a few hours, often fully automatically, thus enabling individualized simulation of electric fields prior to tES intervention.

\yy{In the following sections, we review the formulation and evolution of forward modeling approaches for personalized tES. Section 2.1 covers tissue segmentation, electrode alignment, and mesh generation. Section 2.2 introduces the finite element method for simulating electric fields. Section 2.3 summarizes approaches for simulating electric field distributions induced by tES. Section 2.4 highlights software platforms that enable individualized head model construction. This section provides a theoretical and technical foundation for precise modeling and application of personalized tES.}

\subsection{Tissue Segmentation, Electrode Co-registration, Mesh Generation and Numerical Methods}

% tissue segmentation
The construction of computational head models begins with tissue segmentation, a process that extracts anatomical structures from medical imaging data such as MRI or computed tomography (CT) scans (Fig.\ref{fig:framework} A).

Segmentation involves delineating key tissues relevant to electrical conductivity, including the scalp, skull, cerebrospinal fluid (CSF), gray matter (GM), and white matter (WM). Conductivity values derived from empirical measurements are assigned to segmented tissues to enable electric field distribution calculations (Fig.~\ref{fig:framework} B).% conductivity assignment
Accurate segmentation is the first critical step toward effective modeling.
Segmentation quality is commonly assessed using metrics such as the Dice coefficient and the modified Hausdorff distance, each providing complementary insights. The Dice coefficient measures the similarity between two sets, by computing the ratio of twice their intersection to the sum of their cardinalities. While the Dice coefficient effectively captures the overlap of large voxel clusters, it may overlook subtle discrepancies along region boundaries. To address this limitation, the modified Hausdorff distance~\cite{dubuisson1994modified} is often employed. It calculates the average of two directed Hausdorff distances, where each directed distance represents the maximum deviation from a point in one set to its nearest neighbor in the other. By focusing on boundary dissimilarities, the modified Hausdorff distance offers a finer-grained evaluation of segmentation accuracy. In practice, better segmentation results are indicated by a higher Dice coefficient and a lower modified Hausdorff distance.

% electrode co-registration, mention where?
Following segmentation, electrode co-registration is performed to accurately align stimulation electrodes with the individual's anatomical structures. This step ensures that electrode positions correspond to the same spatial coordinates as the segmented tissues. The procedure is analogous to electrode registration techniques used in EEG studies, typically employing standard scalp landmarks to maintain consistent positioning relative to brain anatomy.

% model construction
The next step is mesh generation for model construction, where the segmented head tissues are discretized into a network of nodes and elements to enable numerical simulations (Fig.~\ref{fig:framework} C). High-quality meshes are critical for accurately modeling electric field distributions and optimizing computational efficiency. The mesh resolution directly affects the model’s ability to capture the complex geometry of the head and faithfully replicate the spatial patterns of electric fields during tES.
The choice of meshing methods depends on the numerical methods employed. In computational modeling, a mesh is a crucial bridge connecting physical problems to mathematical models. Different numerical methods have distinct requirements and approaches for meshing.

The Finite Element Method (FEM) is the most widely adopted~\cite{miranda2006modeling,datta2009gyri,khalil1977parametric, wagner2004three,rampersad2014simulating}, owing to its flexibility in handling complex geometries and heterogeneous, anisotropic conductivity distributions~\cite{wagner2007transcranial}. FEM discretizes the volume into small elements, enabling high-fidelity approximations of electric fields in intricate anatomical structures.
An alternative approach is the Boundary Element Method (BEM), which simplifies the problem by meshing only the boundaries between tissues~\cite{salinas20093d, susnjara2019stochastic,vsuvsnjara2022uncertainty}. By reducing the problem dimensionality, BEM substantially lowers computational demands. However, its applicability is limited in cases involving significant conductivity inhomogeneity or anisotropy~\cite{dayarian2023evaluating, vsuvsnjara2021stochastic}. Despite these challenges, recent methodological advances have improved BEM's robustness for tES applications~\cite{susnjara2019stochastic, vsuvsnjara2022uncertainty}.
\yy{The Finite Difference Method (FDM) presents another option, discretizing the domain on structured grids. While FDM is conceptually simple and computationally efficient for regular geometries, it may face limitations when applied to irregular anatomical boundaries and spatially varying conductivities~\cite{turovets20143d}. However, modern numerical (e.g., MagStim) schemes and some commercial software have implemented strategies that mitigate these challenges, for example through refined discretization and dedicated treatments of non-rectangular geometries. Consequently, FDM-based solvers can still provide accurate solutions for many practical head models.}

\subsection{Finite Element Method for Electric Field Simulation}

% field simulation
The FEM, due to its flexibility, is well-suited for handling complex geometries and is most commonly used for modeling and solving the electric field distribution within the brain. Here, we will use the FEM as an example for a detailed introduction.
Once the whole head mesh model is generated, the accuracy of these simulations critically depends on the quality of the preceding tissue segmentation and mesh generation steps (Fig.~\ref{fig:framework} D).
The meshed head model incorporates the spatial distribution of conductivity across different tissues, and the electric field simulation is governed by Laplace's equation (Fig.~\ref{fig:framework} E). Assuming the absence of net current sources or sinks within the volume, the divergence of the current density ($J$) vanishes. When external currents are applied through electrodes placed on the scalp, the resulting potential distribution ($V$) within the head can be computed by solving the following form of Laplace’s equation~\cite{dmochowski2011optimized}:
\begin{equation}
\label{eq:Laplace1}
\nabla \cdot J = \nabla \cdot ( \sigma E) = - \nabla \cdot \left( {\sigma \nabla V} \right) = 0,
\end{equation}
where $\sigma$ denotes the electrical conductivity of the tissue.
Conductivity is typically assumed to be isotropic—identical in all directions—but can be extended to anisotropic models, as discussed in Section~4.

A unique solution to Laplace’s equation requires appropriate boundary conditions. In tES simulations, both Neumann and Dirichlet boundary conditions are commonly employed. For instance, the potential from the electrodes, denoted as $V_0$, which is determined by factors such as the electrode material, size, and current source, can be applied using the Dirichlet boundary condition:
\begin{equation}
\label{eq:Laplace2}
V = V_0,
\end{equation}
After assembling the global matrix and applying the boundary conditions, the system is solved numerically to obtain the distribution of the electric field.
To efficiently compute the potential distribution, open-source solvers such as \texttt{getDP}~\cite{dular1998general} can be utilized.

\subsection{Simulating Electric Field distribution induced by tES}

After obtaining the electric field in a single direction by anode and cathode, according to \cite{saturnino2019accessibility}, the averaged electric field norm of tDCS can be formulated as Eq.~\eqref{eq:tACS-intensity}.
\begin{equation} 
E = \sqrt{w_xE_x^{2}+w_yE_y^{2}+w_zE_z^{2}},
\label{eq:tACS-intensity}
\end{equation}
where $x$, $y$ and $z$ represent three orthogonal directions, and $w_x$, $w_y$, and $w_z$ denote the weights assigned to the respective orthogonal directions. And because tACS usually has low frequency, a quasi-static approximation can be employed to simulate it. In most cases, we can derive the electric field at peak currents from computational models. The temporal variations of the current merely scale the field without altering its distribution within the brain. When multiple channels are out of phase, simulations should be based on temporal dynamics. Saturnino et al. modeled several in-phase and anti-phase tACS scenarios, finding that they produced complex differences in temporospatial stimulation patterns~\cite{saturnino2017target}.

\yy{tTIS employs a distinct mechanism to stimulate the brain. Two-pair tTIS utilizes two pairs of electrodes, while HD-tTIS employs an array of electrodes for stimulation \cite{huang2020optimization}.  Both types of stimulation are designed to stimulate the brain by the low frequency envelope formed by high frequency current. The effect of tTIS can be evaluated by the modulation depth, which is defined as the difference between the maximal and minimal value of the envelope. The envelope field with the desired direction can be calculated by the formula proposed by Grossman et al.~\cite{grossman2017noninvasive}, shown in Eq.~\eqref{eq:ti2}.}
\begin{equation}
|\vec{E}(\vec{r})|= || (\vec{E}_{1}(\vec{r}) + \vec{E}_{2}(\vec{r})) \cdot (\vec{n}) |
- | (\vec{E}_{1}(\vec{r}) - \vec{E}_{2}(\vec{r})) \cdot (\vec{n}) ||,
\label{eq:ti2}
\end{equation}
where $\vec{E}_{1}(\vec{r})$ and $\vec{E}_{2}(\vec{r})$ are the first and second distinct fields at the location $\vec{r}$ and $\vec{n}$ is an unit vector along the desired direction. 
And Huang et al. provided a comprehensive analysis of the equations governing various forms of tTIS~\cite{huang2020optimization, huang2019can}. They showed the formulae for the modulation depth along the maximum direction.
It is straightforward to find that Eq.~\eqref{eq:ti2} is maximized if and only if $\vec{n}$ falls onto the plane spanned by $\vec{E}_{1}$ and $\vec{E}_{2}$.
Under such premise, the maximal modulation depth equation can be formulated as follows.
\begin{equation}
% M_{\text{max}} = \max_{n} |\vec{E}| = 2 \max_{\alpha} \min(||\vec{E}_{1}|||\cos\alpha|, ||\vec{E}_{2}|||\cos(\alpha - \phi)|)
|\vec{E}| = 2 \max_{\alpha} \min(||\vec{E}_{1}|||\cos\alpha|, ||\vec{E}_{2}|||\cos(\alpha - \phi)|)
\label{eq:ti3}
\end{equation}
where $\phi$ is the angle spanned by $\vec{E}_{1}$ and $\vec{E}_{2}$, and $\alpha$ is the angle between $\vec{E}_{1}$ and the projection of $\vec{n}$ on that plane. 1D-search over $\alpha \in [0, 2\pi)$ can be performed to find the maximal modulation where $\vec{E}_{1}$, $\vec{E}_{1}$ and $\phi$ are known from computational model.
In addition, they proved by the equations that a conventional TES approach (tACS and tDCS) can always be made to reach stronger modulation depth than tTIS regardless of which field orientation one considers.

\subsection{Software for Constructing Personalized Head Models} 

General-purpose FEM software, such as COMSOL Multiphysics (COMSOL, Inc., Burlington, MA), can be utilized for constructing personalized head models. These platforms offer flexible environments for steady-state electrical current simulations, featuring user-friendly interfaces that allow users to define material properties, boundary conditions, and solver settings. Once the model is configured with appropriate tissue conductivities and boundary specifications, the underlying physical equations can be efficiently solved.

\yy{Beyond these general FEM platforms, numerous specialized toolboxes have been developed to streamline various stages of brain stimulation modeling (Table~\ref{tab:forward&optim}). Some, such as FreeSurfer~\cite{fischl2012freesurfer}, Brainstorm~\cite{Tadel2011BrainstormAU}, DUNEuro~\cite{Medani2023BrainstormDUNEuroAI}, Iso2mesh~\cite{Tran2020ImprovingMF}, SCIRun~\cite{dannhauer2012pipeline}, COMETS~\cite{jung2013comets}, %SimBio~\cite{jung2013comets} ,
and Neurophet tES LAB~\cite{karabanov2019can}, primarily focus on forward modeling tasks, including anatomical segmentation, mesh generation, and electric field computation. Others, including SimNIBS~\cite{saturnino2019simnibs} and ROAST~\cite{huang2019realistic}, provide end-to-end frameworks that integrate both modeling and optimization within user-friendly environments.
Collectively, these tools form a comprehensive computational ecosystem supporting individualized tES research. In the following sections, we introduce the computational frameworks underpinning the major tools listed above, and provide a comparative analysis highlighting their respective strengths, limitations, and suitable application scenarios.}

\textbf{SimNIBS}:
\yy{SimNIBS is an open-source software package for head modeling in tES, supporting both forward simulations and the optimization of stimulation parameters. It enables the construction of individualized head models from MRI data and performs electric field simulations using the FEM. Its strengths include user-friendly interfaces, robust segmentation pipelines, and flexibility in handling multi-channel electrode configurations; however, its performance depends on high-quality MRI data and sufficient computational resources, particularly for complex models.}

\textbf{ROAST}:
\yy{
ROAST is an open-source toolbox for personalized head modeling and electric field simulation in MATLAB environments. It leverages SPM12 for tissue segmentation and Iso2mesh for tetrahedral mesh generation, providing anatomically informed modeling of GM, WM, and CSF. ROAST offers a streamlined workflow and supports the processing of MRI data; however, its segmentation accuracy for thin tissue layers can be limited in certain regions.}

\textbf{FreeSurfer}:
\yy{FreeSurfer is a widely used neuroimaging software for cortical surface reconstruction and volumetric segmentation. It provides detailed anatomical labeling of GM and WM, facilitating high-resolution anatomical models suitable for forward modeling of brain activity or electric fields. While FreeSurfer excels in anatomical accuracy, it does not natively perform tES simulations or optimization, although its outputs can be integrated into such pipelines.}

\textbf{Brainstorm}:
\yy{Brainstorm is an open-source platform primarily designed for EEG and MEG source modeling and analysis. It supports head model construction and can interface with FEM solvers for electric field simulations. 
Its strengths lie in interactive visualization and data‑analysis workflows, while built‑in support for tES optimisation is limited and high‑resolution FEM modeling can demand considerable computation time.}

\textbf{DUNEuro}:
\yy{%DUNEuro is a finite element toolbox for solving forward problems in EEG, MEG, and tES.
DUNEuro is an open‑source finite‑element toolbox designed primarily for EEG and MEG forward modelling, with emerging support for tES applications.
It provides flexible numerical solvers for complex head geometries and supports integration with custom head models. Its main advantage is high numerical accuracy, though effective use requires familiarity with FEM principles and programming.}

\textbf{Iso2mesh}:
\yy{Iso2mesh is a MATLAB/Octave toolbox for generating tetrahedral and surface meshes from volumetric or surface data. It is compatible with MRI- or CT-derived segmentations, 
facilitating creation of head‑model meshes that can be used in tES simulation workflows.
While flexible and open-source, it is primarily a mesh-generation toolbox and does not itself perform tES simulation or optimization.}

\textbf{SCIRun}:
\yy{SCIRun is a modular problem-solving environment for biomedical simulation and visualization, supporting forward modeling of tES through flexible mesh generation, numerical solvers, and visualization modules. 
It provides consistent, reusable data structures, simplifying interactive visualization and simulation, but its interface is complex.}

\textbf{COMETS}:
\yy{COMETS is a MATLAB-based platform for modeling tES modalities using individualized or template head models. It allows simulation of electric field distributions for multi-electrode montages and supports protocol design. 
Its limitations include no built-in MRI segmentation, Windows-only operation, and lack of anisotropic tissue conductivity modeling.}

\textbf{Neurophet tES LAB}:
\yy{Neurophet tES LAB is a commercial platform for modeling and optimizing tES. It integrates head modeling, electric field simulation, and optimization workflows, enabling personalized stimulation protocols. Its main advantage is end-to-end usability, though access is restricted to licensed users.}

\begin{table}[t]
\centering
\footnotesize
\caption{Software and methods for forward modeling and inverse optimization.}
\label{tab:forward&optim}
\begin{tabular}{lll} 
\toprule
\textbf{Method} & \textbf{Scope} & \textbf{Website} \\
\midrule
SimNIBS~\cite{saturnino2019simnibs}            & Both         & \url{https://simnibs.github.io/simnibs/build/html/index.html} \\
ROAST~\cite{huang2019realistic}                 & Both         & \url{https://github.com/andypotatohy/roast} \\
Brainstorm~\cite{Tadel2011BrainstormAU}         & Modeling     & \url{https://neuroimage.usc.edu/brainstorm/} \\
DUNEuro~\cite{Medani2023BrainstormDUNEuroAI}    & Modeling     & \url{https://www.medizin.uni-muenster.de/duneuro/forschung.html} \\
COMSOL~\cite{multiphysics1998introduction}      & Modeling     & \url{https://www.comsol.com} \\
Genetic Algorithm~\cite{stoupis2022non}         & Optimization & \url{https://zenodo.org/records/5907212} \\
MOVEA~\cite{wang2022multi}                      & Optimization & \url{https://github.com/ncclab-sustech/MOVEA} \\
\bottomrule
\end{tabular}
\end{table}

\section{Inverse Optimization for personalized tES}

Once the volume-conduction model has been established, a central issue is inter-individual variability in tES delivery. In tDCS, anatomical differences can cause substantial variation in the induced brain electric field even when identical currents are applied~\cite{laakso2015inter}. Personalized dosing helps reduce variability in target-region field strength~\cite{evans2020dose}, but anatomical heterogeneity also complicates the search for optimal electrode configurations~\cite{kasten2019integrating}.

For tTIS, relative focality may be achievable, although its extent in humans remains debated. Evidence suggests that apparent focality occurs chiefly when the region of maximal amplitude modulation of the two interfering currents is also the only region effectively modulated~\cite{violante2023non,rodgers2025increasing,wessel2023noninvasive}. Moreover, tTIS introduces additional design complexity due to its non-convex properties~\cite{grossman2017noninvasive,violante2022non,wang2022multi,huang2019can,huang2018optimized}. In both tDCS and tTIS, electrode geometry—including placement, size, and spacing—can substantially amplify target-region field strength and specificity~\cite{caulfield2021optimizing} (Fig.~\ref{fig:framework}F–I).

This section provides an overview of inverse optimization strategies for personalized tES. Section 3.1 introduces the mathematical formulation of the optimization problem. Section 3.2 presents representative algorithmic approaches for optimizing stimulation parameters and configurations.

\subsection{Formulation of tES Optimization Problem}

Forward modeling enables the estimation of electric field distributions under a given electrode configuration. Based on the simulation results, the effectiveness of stimulation at specific targets can be evaluated (Fig.~\ref{fig:framework} F).
We can use algorithms to identify the optimal configuration for specific targets (Fig.~\ref{fig:framework} H). The inputs of the algorithms are objective functions and leadfield matrix. The objective functions incorporate information regarding the target and constraints. The leadfield matrix, or gain matrix, is a matrix that encapsulates the relationship between the stimulation injected at the scalp and the resulting electric fields in the brain~\cite{dmochowski2011optimized}. This matrix is obtained by sequentially applying a unit current through each candidate electrode. Each element of the matrix represents the electric field in the corresponding area under a given stimulation. By leadfield matrix, we don't need to run the forward model for each potential configuration, significantly reducing the time cost. The electric field can be calculated by Eq.~\eqref{eq:field}:
\begin{align}
\label{eq:field}
\begin{array}{cc}
E = s \cdot L 
\end{array}
\end{align}
where $s$ represents the index of the activation electrode and the corresponding current intensity, and $L$ denotes the leadfield matrix. The leadfield matrix typically encompasses three dimensions: the electrical channel, mesh volume, and orthogonal direction.
The goal of optimization is to maximize or minimize the objective functions,  subjecting to constraints, as shown in Eq.~\eqref{eq:objectives}:
\begin{align}
\label{eq:objectives}
\begin{array}{cc}
&\min/\max \quad \text{Objective function} \\
& s.t. \quad \text{ constraints}
\end{array}
\end{align}
Common goals include maximizing the electric field in the target area~\cite{huang2018optimized,guler2016optimization,Dmochowski2011OptimizedMS} and minimizing deviations between the electric field distribution and preset values~\cite{Dmochowski2011OptimizedMS,Dmochowski2016OptimalUO,Salman2016ConcurrencyIE}. Some complex objective functions are discussed in Section~4.
And the constraints are primarily designed to ensure the safety of human participants (Fig.~\ref{fig:framework} G). The total injection current and the maximum individual injection current are denoted as $I_{tot}$ and $I_{ind}$, respectively. For $N$ candidate electrodes, considering the presence of a reference electrode and according to Kirchhoff’s current law, Eq.~\eqref{con1} stipulates that the sum of the absolute values of the current intensities of the candidates and the reference electrode should be less than $2 * I_{tot}$. The constraints, Eq.~\eqref{con2} and Eq.~\eqref{con3}, are designed to prevent discomfort at the individual electrode interface. The values of \( I_{ind} \) and \( I_{tot} \) are typically set to 1 mA and 2 mA, respectively. However, these values can vary depending on the specific experimental design. For example, in some protocols, such as ISP stimulation, the amplitude can reach as high as 7 mA~\cite{voroslakos2018direct}:
\begin{equation}
{g_1}\left( s \right) = \sum\nolimits_n {\left| {{s_n}} \right|} + \left| {\sum\nolimits_n {{s_n}} } \right| \leq 2{I_{tot}}, \label{con1}
\end{equation}
\begin{equation}
{g_2}(s) = \left| {{s_n}} \right| \leq {I_{ind}}, \label{con2}
\end{equation}
\begin{equation}
{g_3}\left( s \right) = \left| {\sum\nolimits_n {{s_n}} } \right| \leq {I_{ind}}, \label{con3}
\end{equation}

Additionally, constraints can be designed to meet various requirements. Ruffini et al. control the number of active electrodes by penalizing unsuitable solutions~\cite{ruffini2014optimization}. Furthermore, Huang et al. limit the power of the electric field outside the target region to manage the trade-offs between maximal intensity and focality~\cite{huang2018optimized}.

The results of inverse optimization can be directly applied to adjust the stimulation montage in tES, enabling personalized stimulation protocols (Fig.~\ref{fig:framework} I).

\subsection{Optimization Algorithms for Personalized tES}

\begin{figure}[thbp]  
	\centering
	\includegraphics[width=0.98\linewidth]{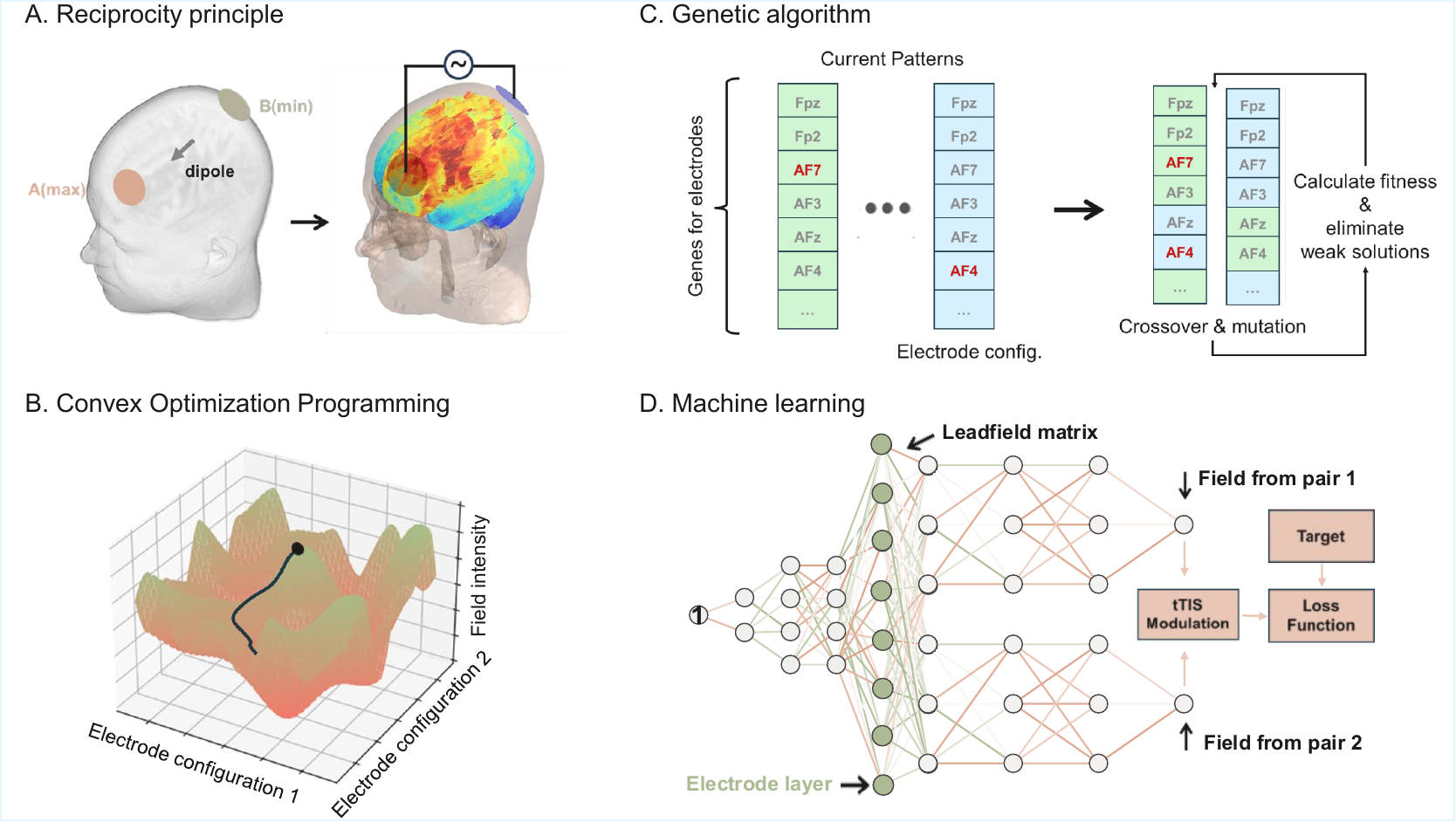}
	\caption{ Schematic diagram of the stimulation optimization methods. (A) Reciprocity principle: The Reciprocity Principle posits that the path of electrical current flow from a stimulation electrode placed on the scalp to a designated area within the brain is equivalent to the path of current flow in the reverse direction, from that specific brain area back to the scalp electrode. Consequently, this principle allows for the utilization of a dipole and a simulation model to determine the optimal electrode configuration for targeted stimulation.
  (B) Convex optimization programming: By iteratively substituting a non-convex problem with solvable convex relaxations or linearized surrogates and refining them, one can leverage convex optimizers to efficiently approach the global optimum of the original formulation.
 (C) Genetic algorithm: This algorithm encodes electrode configurations into the genes of each solution within a population. Genes represent the activation status of the corresponding channel. The solutions represent a stimulation pattern.During each iteration, the solutions undergo crossover and mutation processes at random, resulting in the generation of offspring. Offspring demonstrating superior performance are retained. Ultimately, we can get the optimal solution at the last iteration.
 (D) Deep learning: The approach utilizes a neural network in which a constant unit replaces the input. Fully connected layers are used to generate weights within the electrode layer, which features twice as many nodes as the number of electrodes. Utilizing both the electrode layer and the lead field matrix, the distribution of the electric field and then the loss is calculated. The goal is to optimize the electrode currents that minimize the loss by training the weights of the electrode layer. The panel is adapted from \cite{bahn2023computational}.
	}
	\label{fig:method}
\end{figure}

%exhaustive search
A variety of methods have been proposed for optimizing electrical stimulation strategy. \textbf{The exhaustive search algorithm} is a brute-force method which traverses all possible solutions. Its advantage lies in its universality, suitable for tACS and tTIS, even multi-pair tTIS~\cite{lee2022multipair}, and its ability to obtain the unique, optimal solution of the tES
electrode montage and the corresponding injection current~\cite{lee2020individually,rampersad2019prospects,alekseichuk2019electric,huang2021comparison}. However, given the vast search space, the exhaust search algorithm is  extremely slow and computationally intensive. For example, Rampersad et al. reported that they have to evaluate all 146 million candidates in order to identify the optimal tTIS strategy involving 88 electrodes~\cite{rampersad2019prospects}.	 

% reciprocity principle
\textbf{The reciprocity principle}, first introduced by Helmholtz in 1853, describes the interchangeability of electrical sources and measurement points in a linear, time-invariant system. It states that the effect of a source at a point in the system (e.g., the brain) can be reciprocated if the source and measurement locations are swapped. Leveraging the reciprocity principle (Fig.~\ref{fig:method} A), tES optimization can be approached by first simulating how a source at a target brain location would influence scalp electrodes, and then inverting this relationship to determine optimal electrode placements and activation patterns for effective stimulation of the target\cite{fernandez2016transcranial,cancelli2016simple}.

In practical applications, the lead field matrix encodes this forward information. Based on the reciprocity principle, one can simplify the optimization process by selecting the channel exhibiting maximum intensity to derive an optimal stimulation pattern. Fernandez et al.\cite{fernandez2020unification} further advanced the understanding of the reciprocity principle by presenting a unified framework encompassing related methods. They theoretically demonstrated that Least Squares (LS), Weighted Least Squares (WLS), and reciprocity-based closed-form solutions are all special cases of the extended directional maximization problem. Moreover, LS/WLS and reciprocity-based solutions represent two extremes of the intensity-focality trade-off inherent in this optimization problem. This insight not only validates the efficacy of the reciprocity principle but also provides a means to characterize the performance bounds of other approaches without requiring extensive computational effort\cite{wang2022multi,huang2018optimized}.

%convex optimization  
\textbf{Iterative convex and linear optimization solvers} are widely recognized for their efficiency and effectiveness in addressing complex problems~\cite{dmochowski2011optimized,saturnino2019accessibility,guler2016optimization,ruffini2014optimization}. These approaches typically involve maximizing or minimizing carefully formulated objective functions. Variants of Least Squares (LS) optimization and constrained directional maximization are among the most fundamental and widely adopted techniques in this domain. They form the computational backbone of mainstream tES optimization software, such as SimNIBS~\cite{saturnino2019simnibs, saturnino2021optimizing} and ROAST~\cite{huang2019realistic}, and are crucial for optimizing tTIS~\cite{huang2020optimization,huang2019can}.
For example, a typical optimization problem in tES aims to maximize the electric field intensity at a predefined orientation while satisfying practical constraints, including a total current budget, upper bounds on field strength within avoidance zones, and per-electrode current limits. In this context, Least Squares methods minimize the second-order error between the achieved and desired electric fields, subject to current constraints~\cite{dmochowski2011optimized,fernandez2016transcranial,dmochowski2017optimal}. In contrast, constrained directional maximization seeks to directly maximize the target electric field intensity, also under constraints such as current limits~\cite{dmochowski2011optimized,cancelli2016simple} or avoidance zone regulations~\cite{guler2016optimization,guler2016optimizing}.

When both the objective function and the constraints are convex, powerful frameworks like Disciplined Convex Programming  (Fig.~\ref{fig:method} B) offer effective and scalable solutions~\cite{grant2014cvx}. However, more complex or non-convex scenarios necessitate additional algorithmic strategies. For instance, the integration of \textbf{branch-and-bound algorithms} enables optimizing systems with a limited number of active electrodes—a non-convex problem~\cite{saturnino2019accessibility}. Branch-and-bound methods enhance computational tractability by systematically partitioning the search space and pruning suboptimal regions, ultimately converging to the global optimum. The applicability of convex optimization methods can thus be significantly extended through such hybrid approaches, allowing them to accommodate increasingly complex and realistic constraints.

\textbf{Machine learning} has also been utilized to optimize electrode montages, such as genetic algorithms (Fig.~\ref{fig:method} C) and neural networks (Fig.~\ref{fig:method} D). The genetic algorithm maintains a population of candidate solutions, each solution has a set of proprieties called chromosomes. The genetic algorithm encodes the active state of each electrical channel into the chromosomes of solutions, subsequently engaging in mutation and crossover processes to recombine these chromosomes. With the evaluation of electric field by electrode configurations derived from newly generated solutions, algorithm retains those with superior performance while discarding less effective ones. Through iterative execution, the genetic algorithm converges on the optimal solution. This approach exhibits inherent flexibility in managing non-convex formulations and demonstrates remarkable adaptability in tackling objectives and constraints of increased complexity. For example, Ruffini et al. and Lee et al. employed genetic algorithm to optimize multi-focal stimulation~\cite{ruffini2014optimization, lee2020individually}, and Stoupis et al. used genetic algorithms to achieve focal stimulation over the hippocampus and thalamus using two pairs of tTIS electrodes~\cite{stoupis2022non}. 

Additionally, neural networks have also been employed to address this problem. Bahn et al. applied unsupervised neural networks to optimize tES and achieve multi-target stimulation based on HD-tTIS and HD-tDCS~\cite{bahn2023computational}. In this unsupervised neural network, there is no traditional input. Instead, a constant value is propagated through fully connected layers to generate an electrical configuration. Thereafter, a stimulation network calculates the electric field based on this configuration, from which a loss function is derived. Through this mechanism, the neural network optimizes the electrode currents using repeated backpropagation to achieve field aligned with the target.

Achieving \textbf{focused stimulation} is a key objective of optimization. However, there is an inherent trade-off between focality and intensity in tES. A stronger intensity within the target often results in increased activation outside the target, particularly for deep brain targets. This trade-off can be managed through additional design strategies. The Weighted Least Squares (Weighted-LS) method can enhance the focus on intensity within the target area by applying a weight matrix~\cite{dmochowski2011optimized}. For constrained directional maximization, implementing an extra constraint to limit power outside the target area offers a means to balance these trade-offs~\cite{huang2018optimized}. 
The basic reciprocity-based method primarily focuses on maximizing stimulation intensity without explicit consideration of focality. Dmochowski et al.~\cite{dmochowski2017optimal} addressed this limitation by incorporating L1-constraints, demonstrating that naive maximization strategies can yield suboptimal stimulation patterns. By aligning spatially decorrelated scalp potentials with optimized tES current patterns, they achieved enhanced stimulation efficacy, highlighting the importance of considering spatial focality alongside intensity.

Multi-objective optimization via evolutionary algorithm (MOVEA) is another natural solution for addressing the focused stimulation. Wang et al. provide a comprehensive comparison of various modulation methods and objectives based on the Pareto front~\cite{wang2022multi}. This Pareto front consists of optimal solutions that meet various requirements while respecting trade-off relationships between conflicting objectives such as intensity and focality. And MOVEA is versatile and suitable for both tACS and tTIS based on high definition and two-pair systems.
Furthermore, most algorithms require the definition of a preferred direction, which may not always be available due to prior knowledge or required data. An alternative approach is to optimize the field strength.  MOVEA has proven to be effective in solving it, although this introduces greater complexity due to the need to calculate the norm.

\section{Enhancing tES Modeling through Multimodal Images, Neural Mass Model and Brain-State Feedback}

Although computational models hold great promise for personalized tES, a substantial gap remains between current modeling approaches and real-world neuromodulation outcomes. Bridging this gap requires the integration of additional information to dynamically refine models and enhance their ability to regulate brain activity effectively.
Model updating primarily involves three aspects.
First, anatomical segmentation and anisotropic properties in the head model can be improved using multimodal neuroimaging data (e.g., MRI, DTI). In contrast, model parameters such as tissue conductivity can be adjusted based on recorded stimulation responses (e.g., EEG, fMRI). These refinements increase the accuracy of electric field simulations and improve model adaptability to individual-specific anatomical and physiological characteristics.
\kx{Second, macroscopic models can be integrated with mathematical formulations to enable multi-scale coupling of neural dynamics. By incorporating individual structural connectivity information, personalized neural dynamic models can be linked to anatomical models, capturing the real-time brain responses to external stimulation and inter-individual variability in outcomes.}
Third, the objective functions of the inverse optimization problem can be revised to reflect brain-state-dependent targets, guided by real-time EEG or fMRI data. Previous studies have shown that the effects of tES are highly contingent on the current functional state of the brain~\cite{Krause2023TranscranialES,luo2025frequency}. By leveraging neural signals to infer ongoing brain states, the optimization targets can be dynamically updated, enabling more precise, adaptive, and context-aware neuromodulation.

\kx{In this section, we outline three directions toward enhancing the fidelity and functional relevance of computational modeling for personalized tES. Section 4.1 discussed improvements in anatomical modeling. Section 4.2 focused on linking macroscopic anatomical models with neural dynamic formulations. Section 4.3 introduced adaptive optimization frameworks guided by real-time neural signals, where brain-state-dependent objectives dynamically adjust stimulation parameters for more precise and context-aware neuromodulation. Collectively, these advances outline a pathway toward closed-loop, data-driven, and truly personalized tES modeling and control.}

% \kx{This section encompasses 17 studies published between 2012 and 2025. Among them, 12 introduce pioneering frameworks that address aspects such as structural segmentation, tissue conductivity optimization, multimodal neural activity coupling, and personalized brain dynamics modeling. The remaining five experimental studies focus on integrating individualized brain functional modulation with the characterization of neural dynamics.}

\subsection{Refining Forward Models through Segmentation, Conductivity, and Anisotropy Updates}

% \kx{segmentation through ML?}

\begin{figure}[thbp]  
	\centering
	\includegraphics[width=0.98\linewidth]{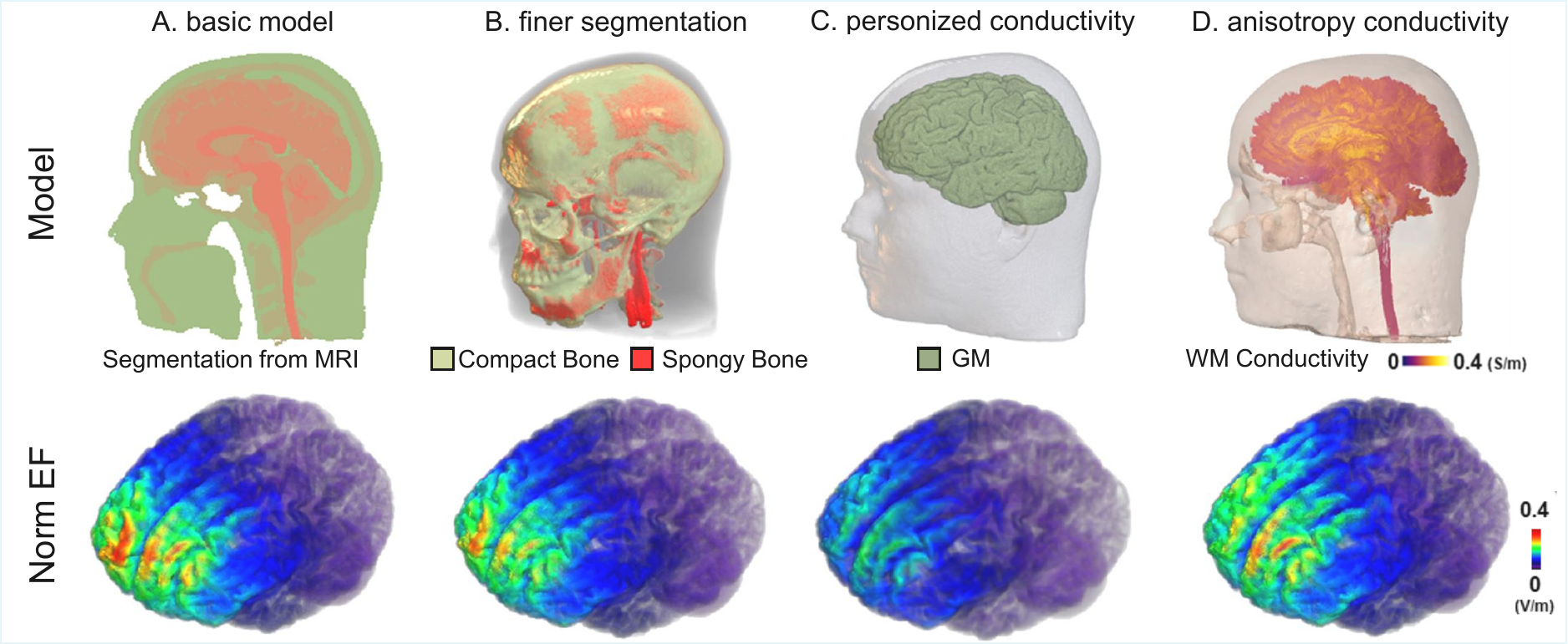}
	\caption{The influence of model updates on the norm of the electric field.
(A) The Standard Model: The upper panel displays a segmentation with five tissue types, each assigned conductivities based on standard values from prior research (WM: 0.126; GM: 0.275; CSF: 1.654; Bone: 0.01; Scalp: 0.465). The lower panel illustrates the distribution of the electric field's magnitude under a 1 mA tDCS from F3 to F4 in the EEG10-10 system. Following visualizations follow the same settings.
(B) Bone subdivisions. The bone is segmented into two distinct categories: compact bone with a conductivity of 0.008 and spongy bone with a conductivity of 0.025.
(C) Conductivity optimization. The conductivity of Gray Matter is adjusted to its highest reported value in previous research, being 0.6. In practice, conductivity optimization can be combined with sEEG to fine-tune the model's parameters.
(D) Anisotropy model. The upper panel depicts the conductivity of White Matter. The conductivity of tissues is configured as anisotropic by direct mapping based on linear rescaling of tensors from diffusion MRI. The upper threshold for conductivity is established at 2 S/m, with a maximum ratio of 10 between the largest and smallest conductivity eigenvalues. 
The unit of conductivity: S/m. The unit of electric field: V/m. Abbr: WM, White Matter; GM, Gray Matter; CSF, Cerebrospinal Fluid; EF, electric field; tDCS, transcranial Direct Current Stimulation.
	}
	\label{fig:update}
\end{figure}

A substantial discrepancy often exists between the electric fields predicted by computational models and those empirically measured (Fig.~\ref{fig:update}). This mismatch is largely attributable to individual variability in tissue properties and limitations in anatomical segmentation.
Improving segmentation fidelity, both in terms of accuracy and anatomical detail, is crucial for enhancing forward modeling precision~\cite{puonti2020accurate, jalal2021robust} (Fig.~\ref{fig:update} B).
% For instance, Weise et al.~\cite{weise2022effect} demonstrated that the inclusion of the meninges, often omitted due to computational cost, can significantly alter the peak electric field intensity.
\kx{For instance, Weise et al.~\cite{weise2022effect} demonstrated that the inclusion of the meninges, often omitted due to computational cost, can significantly alter the peak electric field intensity. Using a boundary element-fast multipole method to incorporate the meninges into electric field simulations, they found that the cortical electric field strength increased by approximately 30\% during tES when the meninges were included.}
\kx{Similarly, Wang et al.~\cite{wang2021influence} reported that in tTIS, the Pearson correlation coefficient between electric field distributions derived from layered-skull and single-layer skull models was only 0.746, which highlights substantial discrepancies in predicted electric field distributions across forward models employing different skull representation strategies.}
% \todo{add more about ML}
These findings underscore the importance of incorporating detailed and individualized tissue definitions into the model. This challenge is further compounded in pediatric or neonatal populations, where immature tissue development and limitations in imaging resolution complicate segmentation and conductivity estimation~\cite{despotovic2013realistic}.

\kx{Additionally, a recent trend in development is the integration of deep learning into head model segmentation. It has demonstrated outstanding performance in image segmentation and has been increasingly applied to the delineation of anatomical brain structures.
Rashed et al.~\cite{rashed2019development} introduced ForkNet, which leverages convolutional neural networks (CNNs) for automatic tissue segmentation from MRI data, thereby enhancing both the accuracy and efficiency of head tissue delineation. Hirsch et al.~\cite{hirsch2021segmentation} developed MultiPrior, a model that integrates deep volumetric networks with multiple spatial priors to improve segmentation accuracy and robustness. Stolte et al.~\cite{stolte2024precise} introduced the GRACE model, which effectively segmented personalized head models of elderly individuals, facilitating high-fidelity modeling of age-related neurological conditions.
}

Tissue conductivity is another key issue in modeling electric field distribution~\cite{huang2017measurements}. Conductivity values can vary significantly across individuals, tissue temperature, and even stimulation frequency~\cite{geddes1967specific}. To accurately set the tissue conductivity values, head model parameters can be calibrated using intracranial recordings, such as intracranial EEG (iEEG), by aligning predicted and measured values in localized brain regions (Fig.\ref{fig:update} C). Several techniques have been proposed for estimating and optimizing conductivity, as well as for quantifying the reliability of model predictions. For example, Huang et al.~\cite{huang2017measurements} developed an optimization algorithm that adjusts conductivity values to minimize the discrepancy between modeled and observed iEEG potentials.
\kx{Carvallo et al.~\cite{carvallo2018biophysical} introduced a biophysical modeling framework that utilizes stereoelectroencephalography (sEEG) data to estimate tissue-specific conductivity parameters by comparing simulated and recorded potential distributions, thereby significantly improving the accuracy of individualized conductivity estimation across brain tissues.
Saturnino et al.~\cite{saturnino2019principled} developed a systematic uncertainty analysis framework to assess how variability in tissue conductance affects electric field computations, offering a theoretical foundation for optimizing conductance parameters and calibrating individualized head models.}
This calibration can enhance field estimation accuracy across the brain. 

% \todo{more specific for tissue conductivity}
Most conventional head models simplify conductivity as a scalar value, without considering the anisotropy of conductivity. However, this assumption overlooks the directional complexity of neural tissue. The presence of aligned white matter tracts gives rise to anisotropic conductivity, which significantly influences the spatial distribution of the electric field~\cite{Tuch2001ConductivityTM, Opitz2011HowTB} (Fig.\ref{fig:update} D). Suh et al.~\cite{Suh2012InfluenceOA} demonstrated that incorporating anisotropic skull conductivity can reduce off-target electric fields by 12-14\%, thus enhancing spatial specificity. A commonly used approach for modeling anisotropy involves constructing conductivity tensors based on diffusion MRI-derived diffusion tensors and fractional anisotropy metrics. These tensors are rescaled according to known tissue conductivity values and integrated into FEM frameworks. This enables more biologically realistic simulations that account for directional dependencies in current propagation.

% \subsection{neural mass modeling \todo{xx}}

\subsection{Coupling from Electric Field Distribution to Neural Activity}

% NMM intro
\kx{Beyond improving physical-level accuracy, another promising direction is to incorporate individualized neural dynamics into the model~\cite{deco2011emerging, breakspear2017dynamic}. Neural dynamics shape how external stimulation modulates the brain’s functional responses. Therefore, incorporating neural population dynamics into individualized modeling is essential for enhancing the physiological realism of computational models~\cite{cottone2018new}.
A widely used framework is the Neural Mass Model (NMM)~\cite{jansen1995electroencephalogram, david2003neural}. NMM describes the average activity of excitatory and inhibitory neuronal populations within cortical columns by employing a nonlinear mapping between the mean membrane potential and firing rate of neuron groups.
% These interactions and feedback loops are represented through a set of ordinary differential equations.
Compared with single-cell models, the NMM offers a balance between biophysical interpretability and computational efficiency, capturing macroscale oscillations, rhythmic synchronization, and dynamic responses to external stimulation. Consequently, it is widely employed for generating and interpreting EEG, MEG, and fMRI signals.
}

% NMM modeling
\kx{Neural mass models represent neural populations as fundamental units, forming local dynamical modules based on input–output relationships between excitatory and inhibitory subpopulations. In personalized modeling, researchers commonly use structural connectivity derived from imaging data (e.g., DTI) to determine the coupling matrix between nodes in the NMM. Model parameters are subsequently optimized against individual EEG or fMRI data to reproduce subject-specific neural dynamics~\cite{babajani2006integrated, sanz2015mathematical, cakan2023neurolib}. This structure-function integrated modeling framework enables the NMM to serve as the dynamic core of an individualized head model, allowing for predictions of neural responses and state transitions under various stimulation patterns.}

% NMM application
\kx{In tES research, integrating NMM with electric field modeling offers a novel framework to understand how electrical stimulation regulates neural dynamics. Traditional volumetric conductor models treat the brain as a linear conductor, focusing on the distribution of the electric field. However, these models overlook the nonlinear and state-dependent characteristics of neural responses under external electric fields. In contrast, NMMs capture the mean activity of excitatory and inhibitory populations within cortical regions by deriving population-level equations from spiking neuron networks. This framework enables biologically realistic simulations of how external stimulation modulates local oscillatory rhythms, interregional synchrony, and large-scale network coupling~\cite{ali2013transcranial, neuling2012good}. Kunze et al.~\cite{kunze2016transcranial} integrated FEM with NMM to simulate large-scale networks, demonstrating that tDCS modulates resting-state connectivity and network dynamics.}

% \todo{NMM IN TES}

\kx{Embedding NMM within stimulated regions of personalized head models allows simulation of functional-level neurodynamic responses.
% One feasible approach is to input electric field distributions from FEM calculations into NMM equations to examine how stimulation modulates population-level oscillations.
Karimi et al.~\cite{karimi2025precision} developed an NMM-based framework that combines MRI/DWI-derived connectivity with electric field simulations, enabling the optimization of individualized stimulation parameters. This approach enables the optimization of stimulation parameters to target specific regions while precisely regulating the whole-brain dynamics. Future advancements lie in coupling NMM with multi-scale neuronal models. For instance, Chung et al.~\cite{chung2022key} integrated FEM with single-neuron models to uncover how tES effects depend on neuron type, field orientation, and connectivity. Integrating neural dynamics into personalized models can capture individual variability in cortical excitability, oscillatory states, and connectivity. This approach facilitates the accurate prediction of stimulation effects and the optimization of parameters tailored to individual neural states.}

\subsection{Incorporating Brain-State Dynamics into Inverse Optimization}

Conventional inverse optimization approaches in tES often treat the brain as a passive electrical medium, neglecting its intrinsic nonlinear and state-dependent behavior. However, accumulating evidence suggests that the efficacy of neuromodulation is highly dependent on the brain's dynamic state at the time of stimulation~\cite{Krause2023TranscranialES,Frhlich2010EndogenousEF}. These brain states—reflected in EEG rhythms, BOLD fluctuations, or patterns of functional connectivity—can critically influence the brain’s responsiveness to external stimulation. Accordingly, integrating dynamic brain-state information into the optimization framework represents a key step toward adaptive and truly personalized tES.

%\subsubsection{State-Aware Objective Functions}
One promising direction is to embed brain-state features directly into the objective function of the optimization framework. Rather than minimizing electric field errors in anatomical space alone, state-aware optimization seeks to maximize neuromodulatory efficacy by targeting functionally relevant network states. For example, network controllability theory offers a powerful lens for understanding how exogenous inputs can drive neural systems from one state to another. In this framework, optimal stimulation parameters can be derived by minimizing the energy required to transition the system toward a desirable brain state, given the individual's current functional configuration. 

The first step in brain-state-informed optimization is the acquisition of relevant neural signals during tES. EEG is the most widely used modality to assess tES-induced neural changes before and after stimulation~\cite{Wang2023TheEO,Tashiro2020ProbingEA,Mulyana2022OnlineCR}. However, accurately capturing brain activity during stimulation remains challenging due to strong electrical artifacts. To address this, alternative modalities such as fMRI and sEEG have been employed for simultaneous recording with tES~\cite{louviot2022transcranial, Mulyana2022OnlineCR, GhobadiAzbari2020fMRIAT, Mulyana2021OnlineCR, Lou2024ADF}. These studies have not only demonstrated the efficacy of tES but also revealed significant discrepancies between simulated and empirically observed neural responses~\cite{louviot2022transcranial, luo2025frequency}, highlighting the influence of individual variability and dynamic brain states.
One explanation for these discrepancies is the oversimplification of the brain as a linear conductor in computational models. In reality, the brain is a complex, nonlinear, and adaptive system. Recent studies have shown that tES can modulate brain networks rather than isolated regions~\cite{saturnino2021optimizing, Ruffini2014OptimizationOM, Ali2013TranscranialAC}. For example, Rostami et al. demonstrated that tACS modulates cortical connectivity within the dorsal attention network, as evidenced by changes in phase-locking values~\cite{Rostami20206HT}. Mulyana et al. further advanced this direction by proposing a closed-loop tES-fMRI framework to optimize individualized stimulation of frontoparietal networks, resulting in enhanced functional connectivity and improved working memory~\cite{Mulyana2022OnlineCR}. Krause et al. suggested that individual variability in neuronal phase preference contributes to heterogeneous tES outcomes, which could be better explained using oscillator-based models~\cite{Krause2023TranscranialES, Krause2019TranscranialAC}.

To better capture the temporal evolution of brain activity, hidden Markov models (HMMs) have been widely employed. For example, Kasten et al. used HMMs to analyze MEG data collected before and after tACS, revealing that spontaneous brain states and their underlying functional networks vary in susceptibility to stimulation~\cite{Kasten2022TheHB}. Extending this approach, Brown et al. applied high-definition alpha-frequency tACS during a continuous performance task while recording fMRI data. Their results demonstrated that stimulation modulated the temporal dynamics of brain states, as identified by HMMs~\cite{Brown2023TranscranialSO}.
We can design objective functions based on brain states and obtain corresponding stimulation strategies. Combining the method with control theory, the brain can be modeled as a dynamic system influenced by external input. A simplified representation is given by a discrete-time linear time-invariant system:

\begin{equation}
    x(t + 1) = Ax(t) + Bu(t),
    \label{dynamic}
\end{equation}

where vector \( x(t) \in \mathbb{R}^N \) denotes the state of the brain network with \( N \) regions at time \( t \). The matrix \( A \in \mathbb{R}^{N \times N} \) describes the intrinsic connectivity among brain regions, representing the network topology. The matrix \( B \) captures the distribution of external input, and vector \( u(t) \in \mathbb{R}^N \) represents the control signal applied at time \( t \).

Control strategies can be broadly categorized as either model-driven or data-driven. In model-driven control, matrix \( A \) is usually derived from prior knowledge, such as structural (white matter tracts), functional (statistical dependencies), or effective (causal interactions) connectivity~\cite{gu2015controllability,tanner2024multi,luo2025mapping}. The leadfield matrix from electrical modeling can serve as matrix \( B \), quantifying how each electrode influences brain regions.
In contrast, data-driven approaches aim to estimate both \( A \) and \( B \) directly from empirical neural recordings. Given time-series data that reflect evolving brain states \( x(t) \), the goal is to design optimal control inputs \( u(t) \) that drive the system toward desirable cognitive or clinical outcomes.

\section{Limitations and Future Directions}

We have introduced simulation and optimization techniques for personalized tES. Computational modeling has proven crucial in enhancing tES performance. However, some limitations still hinder tES performance, which should be addressed.

To fully harness the advantages of computational models, their integration with real-world stimulation hardware is essential. However, this integration is complicated by the diversity of stimulation devices and experimental environments. Most models are built on idealized assumptions, such as standardized equipment configurations and accurate parameter settings, which may not reflect real-world conditions.
In practice, factors such as head curvature and subject motion during experiments can result in uneven contact pressure between electrodes and the scalp. This may displace electrodes or alter contact quality, ultimately distorting the electric field distribution and reducing stimulation efficacy~\cite{iyer2005safety}. Although methods have been developed for real-time electrode localization~\cite{Clausner2017PhotogrammetryBasedHD,JaffeDax2019VideobasedMR}, current tES frameworks generally lack the capability to dynamically account for such variabilities and adapt the model accordingly. Recent studies have begun to explore more adaptive and online modeling strategies to address this limitation~\cite{liu2024modelling,liang2022online}.
Additionally, variability in electrode properties—such as size, material, and surface texture—can significantly influence stimulation outcomes~\cite{lopez2014dry,ng2022multi}. Despite its importance, incorporating these properties into optimization algorithms remains challenging, largely due to the absence of well-defined models to quantify their effects. While electrode size has been shown to play a critical role, no universally accepted equations currently exist to capture the influence of other electrode characteristics, such as material composition and geometry.

The analysis of brain signals during tES presents a promising direction, particularly with the emergence of tES devices that support simultaneous neural signal acquisition, thereby enabling the development of closed-loop systems. Among these, EEG is frequently employed alongside tES due to the compatibility of their underlying principles.
However, recording EEG during ongoing tES poses significant technical challenges. The stimulation introduces substantial artifacts, often several orders of magnitude larger than the neural signals of interest, and may even saturate EEG amplifiers~\cite{liu2018immediate, noury2017phase}. In addition to electrical artifacts, physiological noise from sources such as cardiac and respiratory activity further contaminates the EEG recordings~\cite{noury2017phase, noury2016physiological}. As a result, the signal-to-noise ratio (SNR) is drastically reduced, making direct interpretation of EEG during tES infeasible.
To mitigate this issue, denoising techniques have been proposed. Several offline denoising algorithms have been developed for tES-EEG data~\cite{haslacher2021stimulation, kohli2015removal, vosskuhl2020signal}, and more recently, real-time approaches have emerged to facilitate online monitoring of brain activity. For example, Guarnieri et al. introduced an adaptive spatial filtering method, known as alternating current regression (AC-REG), specifically designed to attenuate tACS-induced artifacts in real time~\cite{guarnieri2020computationally}.
Despite these advances, major challenges remain. The most important of these is the lack of ground truth data, which makes it difficult to verify whether the denoised EEG accurately reflects true neural activity or still contains residual artifacts. Moreover, since tES itself modulates brain activity, the denoising process risks removing not only artifacts but also genuine stimulation-induced neural changes. As a result, validating the effectiveness and fidelity of denoised signals remains an open and actively investigated issue~\cite{kohli2020machine}.

%frequency \todo{high-frequency}
The frequency of the injected current is another critical factor that warrants careful consideration in computational modeling. Contemporary models predominantly emphasize the amplitude of stimulation and the resulting electric field distribution, often overlooking the significant influence of current frequency. In the case of tACS, which typically operates at low frequencies, the quasi-static approximation is commonly employed. This approximation models the electric field at peak current amplitudes, under the assumption that temporal variations in current intensity only scale the magnitude of the electric field without affecting its spatial distribution. 
Nevertheless, an increasing number of studies have reported that frequency and phase can substantially affect both the distribution and efficacy of stimulation~\cite{saturnino2017target, wang2021influence}. These effects become especially pronounced at the higher carrier frequencies used in tTIS. For instance, Liu et al. observed a slight but measurable decrease in the field magnitude at temporal interference frequencies exceeding 2 kHz, which may be attributable to frequency-dependent changes in tissue conductivity~\cite{Liu2024TemporalIS, Gabriel1996TheDP}.
Beyond physical field properties, frequency also plays a critical role in shaping neural responses. Variations in stimulation frequency can significantly modulate neuronal excitability and synaptic dynamics, influencing the effectiveness of neuromodulation~\cite{hutcheon2000resonance}. Therefore, it is imperative to expand computational models to incorporate frequency-dependent physiological responses. This can be achieved by integrating large-scale concurrent neural recordings and leveraging advanced signal analysis methods. Such data-driven approaches are essential for developing predictive models that elucidate the complex relationship between stimulation frequency and brain function.

Additionally, the orientation of the electric field is pivotal, as it can dictate the excitatory or inhibitory responses of neurons. Especially for tTIS, which regulated brain by two pairs of electrodes, a given field orientation can significantly enhance the focality of the envelope field. While optimization algorithms have been developed to refine this aspect, there remains a gap in our understanding of how to precisely configure the field's direction for specific targets. In the case of the cerebral cortex, a vertical orientation is commonly employed~\cite{dmochowski2011optimized}. However, for other tissues, defining a more precise target necessitates a deeper integration of neuroscience knowledge. Designing the objective function based on advanced neuroimaging could be a viable approach. For instance, fiber orientation obtained through diffusion MRI could be incorporated into the objective function.

\section{Conclusion}
tES is an effective non-invasive neuromodulation technique for treating neurological diseases and enhancing cognitive performance. Personalized tES implementation is crucial for mitigating variations in electric field distribution resulting from anatomical differences. Forward modeling and inverse optimization are vital stages in the computational modeling of personalized tES, facilitating the refinement of the stimulation strategy. Optimization methods can be utilized to design stimulation strategies that are more focused or multi-targeted. 
\kx{Integrating personalized tES computational modeling with neuroimaging and neurodynamic modeling offers excellent potential for enhancing stimulation precision. However, how the external tES interacts with the internal brain's activity remains a topic of ongoing discussion. Combining tES with multimodal, high-precision personalized computational models is expected to become a key direction for future research.}

\section*{Acknowledgement}
We would like to express our sincere gratitude to Dr. Zhichao Liang and Dr. Xinke Shen for their invaluable suggestions during the preparation of this manuscript.
This work was supported by the National Natural Science Foundation of China (62472206), National Key R\&D Program of China (2025YFC3410000), Shenzhen Science and Technology Innovation Committee (RCYX20231211090405003, KJZD20230923115221044), Guangdong Provincial Key Laboratory of Advanced Biomaterials (2022B1212010003), and the open research fund of the Guangdong Provincial Key Laboratory of Mathematical and Neural Dynamical Systems, the Center for Computational Science and Engineering at Southern University of Science and Technology.

\bibliographystyle{unsrt}
\bibliography{main}

\end{document}